# Ag and N acceptors in ZnO: ab initio study of acceptor pairing, doping efficiency, and the role of hydrogen

O. Volnianska[1], P. Boguslawski[1,2], and E. Kaminska[3]

1. Institute of Physics PAS, al. Lotnikow 32/46, 02-668 Warsaw, Poland
2. Institute of Physics, Kazimierz Wielki University, Powstancow Wielkopolskich 2, 85-064 Bydgoszcz, Poland.
3. Institute of Electron Technology, al. Lotnikow 32/46, 02-668 Warsaw, Poland

*Abstract*.
Efficiency of ZnO doping with Ag and N shallow acceptors, which substitute respectively cations and anions, was investigated. First principles calculations indicate a strong tendency towards formation of nearest neighbor Ag-N pairs and N-Ag-N triangles. Binding of acceptors stems from the formation of quasi-molecular bonds between dopants, and has a universal character in semiconductors. The pairing increases energy levels of impurities, and thus lowers doping efficiency. In the presence of donors, pairing is weaker or even forbidden. However, hydrogen has a tendency to form clusters with Ag and N, which favors the Ag-N aggregation and lowers the acceptor levels of such complexes.

# I. INTRODUCTION

Efficient p-doping of ZnO is a problem that is not satisfactorily solved yet. Among a variety of investigated species, Ag and N lead to particularly good results. Ag doping was studied in Refs. [1-9], and its acceptor character was confirmed. In particular, ZnO:Ag layers grown by sputter deposition are *p*-type with concentrations up to $10^{18}$ cm$^{-3}$ and hole mobilities of about 1 Vm/sec$^2$.[10] Doping ZnO with N also leads to *p*-type conductivity,[11-15] and the achieved parameters are comparable to those of ZnO:Ag. Wei *et al*. [15] have shown that conductivity of as-grown ZnO doped with N is *n*-type, and it transforms to the expected p-type after annealing at 600 ºC. In the as-grown samples, N can preferentially be incorporated as N$_2$ molecules which are donors, and which out diffuse during annealing, while the substitutional N:O remains in the samples.[15] Finally, promising results were obtained with dual acceptor doping, using simultaneously As and N,[16] Ag and N,[17] or P and N.[18] In any case, *p*-doping efficiency of ZnO is low: the measured concentrations of free holes are typically lower than those of the intentional acceptors by at least one or two orders of magnitude. Several aspects of Ag doping were theoretically investigated in Refs. [19-21], where it was concluded that Ag is the most efficient group-I$^A$ acceptor. N doping was investigated in Refs. [22,23]. It was also shown that doping efficiency can be limited by formation of pairs and larger nano-aggregates of few atoms.[20]

Ideally, the concentration of free carriers is determined by the concentration of incorporated impurities and their ionization energy. In practice, several processes limit doping efficiency, such as the compensation by native defects,[24] or incorporation of a dopant at the "wrong", i.e., interstitial or antisite, position. Another process limiting the doping efficiency is the formation of few-atom aggregates, or even nano-inclusions of second phases, which can be of importance at high doping rates used in the ZnO technology. Indeed, for concentrations below the solubility limit it is assumed that the distribution of impurities is random, which typically is correct. However, many impurities reveal the tendency toward clustering, which is expected to take place when the solubility limit is exceeded. Moreover, theoretical calculations show that acceptor-acceptor interactions result in a tendency to form nearest neighbor acceptors pairs. This appears to be a general phenomenon in semiconductors. Indeed, this is the case of e.g. ZnO:Cu[25] (for which the pairing is confirmed by experiment),[26] ZnO:Ag,[20] and other host-impurity systems,[27] with a typical binding energy of a pair $E_{bind}$ of about 0.3 eV. Note that the pairing energy can be increased by the magnetic coupling between transition metal impurities. The impact of formation of N-N nearest neighbor pairs on electrical conductivity in ZnO was analyzed in Ref. [28]. The authors considered the presence of pairs, which are present even when the distribution of N is random. However, the actual pair concentration can be higher because of the acceptor-acceptor coupling that is investigated in this work.

Difficulties with achieving *p*-type ZnO can also be due to the presence of hydrogen in the samples. Hydrogen is known to be a non-intentional donor, typically present at high concentrations, either as a product of growth process or as a result of in-diffusion from atmosphere. Properties of H in ZnO relevant for this paper were discussed in Refs. [29-32], and they are not investigated here. H is a shallow donor [29], and thus it compensates acceptors such as Ag or N becoming a positively charged H$^+$ ion, *i.e.*, a proton. The simultaneous presence of both acceptors and donors in a semiconductor corresponds to the so-called co-doping.[33] Stable sites of H$^+$ in ZnO are bond centers[29,31,32], and hydrogen is mobile with the low diffusion barrier of 0.5 eV.[30]

In this work, we theoretically investigate efficiency of dual doping of ZnO with Ag and N, taking also into consideration the presence of hydrogen in ZnO layers. The calculations are performed using the generalized gradient approximation (GGA) to the density functional theory, and the details are given in Sec. 2. We first analyze energetics of several simple configurations of impurities (pairs, triangles, complexes with H), which are likely to occur in ZnO:(Ag,N), since this is the actual configuration of defects which defines their electronic structure. In Sec. 3 we show that formation of acceptor-acceptor pairs in a crystal is driven by the same mechanism as the formation of, e.g., $N_2$ molecules in vacuum, and it consists in the formation of a molecular-like bond accompanied by the formation of a bonding-antibonding pair of orbitals. Formation of molecular-like bonds between acceptors is predicted to be universal, and to occur in all semiconductors. For similar reasons, formation of triangles is also favorable. A characteristic feature of dual doping with Ag and N is the fact that it involves substitution on both sublattices, as Ag and N substitute Zn and O, respectively. This leads to formation of mixed Ag-N nearest neighbor pairs, with different binding energy, electronic structure *etc.* than those of Ag-Ag and N-N second neighbor pairs. The presence of donors, and in particular of H, strongly affects formation of nano-aggregates, as it is shown in Sec. 4. In Sec. 5 we analyze the impact of Ag-N-H complexes on the electronic structure of ZnO:(Ag,N) and on the doping efficiency. Section 6 summarizes the paper.

## II. METHOD OF CALCULATIONS

Calculations based on the density-functional theory were performed within the generalized gradient approximation,[34,35] using QUANTUM-ESPRESSO code.[36] As it was discussed in Ref. [20], the underestimation of the band gap by GGA has a negligible impact on shallow acceptor states which are both energetically close to the valence bands and derived from valence states. We have employed ultrasoft atomic pseudopotentials,[37] and the plane wave basis with the kinetic energy cutoff of 30 Ry, which provided a good description of II-VI oxides. Orbitals that were chosen as valence orbitals are 3$d$, 4$s$ for Zn, 2$s$, 2$p$ for O, 5$p$, 4$d$, 5$s$, for Ag, 2$s$, 2$p$ for N, and 1$s$ for H. Methfessel-Paxton smearing method with the smearing width of 0.136 eV has been used to account for partial occupancies.[38] Ionic positions were optimized until the forces acting on ions were smaller than 0.02 eV/Å. To study the impurities, large unit cells with 128 atoms were employed, and the Brillouin zone summations were performed using the Monkhorst-Pack scheme with a 2×2×2 $k$-point mesh for the wurtzite structure. To correct for the Coulomb interactions between charged impurities and their images in different supercells, inherent in the supercell method, we used the method based on the Ewald technique elaborated by Tosi [39]. The charges of Ag, N, and H are approximated by localized Gaussian charge distributions with appropriate charge values (e.g., -0.5 $e$ and +1 $e$ for the acceptor and H forming an Ag-N-H complex, respectively, where $e$ is the proton charge) depending on the actual configuration. The calculation showed that the corrections to electrostatic energy are about 0.1 eV. Finally, the binding energy $E_{bind}$ of complexes of substitutional Ag$_{Zn}$, N$_O$ and interstitial hydrogen is defined as the difference in the total energy of the system with isolated defects and the system with aggregated defects. A cluster is stable when $E_{bind}$ is positive.

## III. Ag-N PAIRS

### A. Isolated Ag and N

The calculated energy levels of isolated Ag and N in ZnO are very close. The acceptor-induced triplet level is split into a doublet and a singlet by the hexagonal wurtzite crystal field. The doublet is situated above the singlet, and its energy is 0.17 and 0.16 eV for Ag and N, respectively, while the crystal field splitting of the triplet amounts to 0.06 eV for Ag and 0.04 eV for N. Thus, the calculated impurity levels $E_{imp}$ of both acceptors are practically the same to within our accuracy, and differences in doping efficiencies stem from different formation energies, different dependence of incorporation on the growth conditions, *etc*.

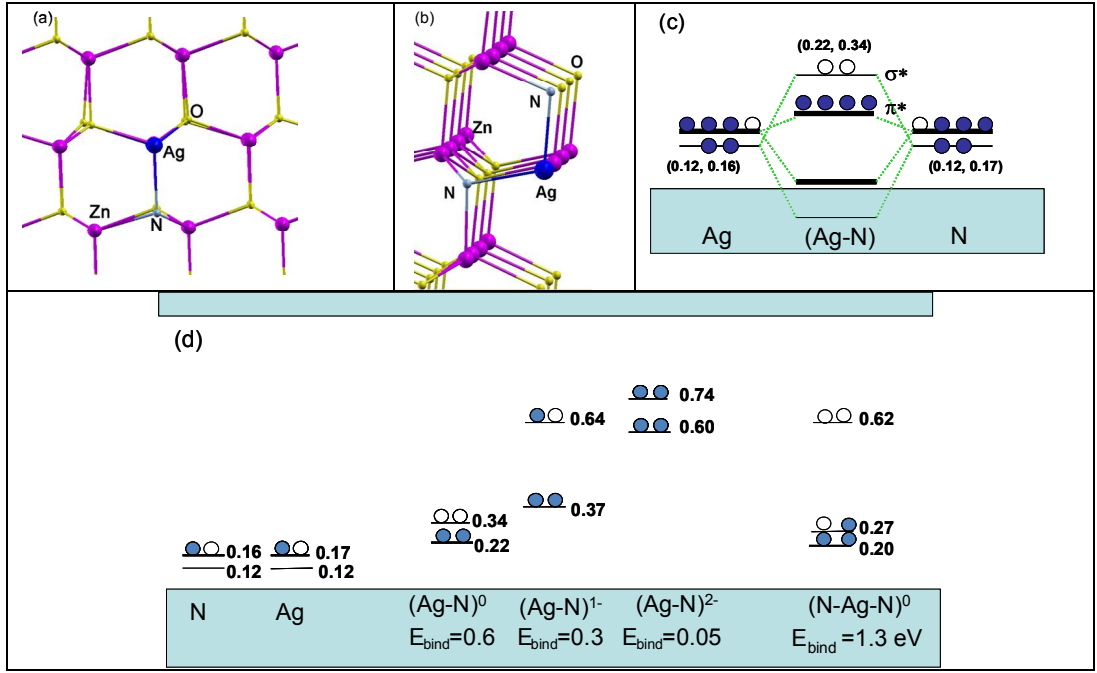

FIG. 1. (Color online) (a) Configuration of an Ag-N nearest-neighbor pair, and (b) of the N-Ag-N nearest-neighbor triangle. Magenta, yellow, big navy blue and small pale blue balls represent Zn, O, Ag and N atoms, respectively. (c) Energy levels relative to the top of the valence band of isolated Ag, N, and of the Ag-N nearest-neighbors pair, with the indicated antibonding σ* and π* combinations. (d) Energy levels isolated Ag, N, Ag-N nearest-neighbor pair in the $q$=0, 1- and 2- charge states, and of a neutral N-Ag-N triangle. The numbers in (c) and (d) give the calculated level energies in eV. $E_{bind}$ is in eV.

### B. Ag-N pairs and molecular-like bonds

Formation of Ag-Ag acceptor pairs and larger nano-aggregates was previously analyzed in Ref. [20], and it was found that Ag-Ag pairing is energetically favorable. The calculated binding energy $E_{bind}$ of a neutral Ag-Ag pair, *i.e.*, the energy gain with respect to the case of isolated dopants, is 0.35 eV.[20] In the case of dual doping with Ag and N, there is a possibility of formation of mixed nearest neighbor pairs $Ag_{Zn}$-$N_O$, in which the Ag-N distance of about 1.87Å is close to the host bond length, and is about twice smaller than that of a $Ag_{Zn}$-$Ag_{Zn}$ or $N_O$-$N_O$ pair, 3.06 Å. The calculated binding energy of a neutral Ag-N pair is 0.62 eV. This value is larger than $E_{bind}$ for Ag-Ag, and it implies the stability of the Ag-N pair at typical growth or annealing temperatures. Both the Ag-Ag and Ag-N coupling is short-range, since *e.g.* the binding of the Ag-Ag second neighbors is about 10 meV. A similar short range character of the binding was previously found for other systems.[20,25,27]

The binding of an acceptor-acceptor pair can be explained within a model of a covalent molecular-like bond. In a molecule, valence atomic orbitals form a bonding and an antibonding combination, and the bonding-antibonding splitting increases with the decreasing interatomic distance. Binding of a molecule occurs only for a partial occupation, when the binding energy originates in the higher occupation of the bonding than of the antibonding states. For this reason, noble gases like He or Ne with filled valence states do not form molecules.

The same effect occurs for a pair of two acceptors. In this case, the formation of a molecular-like bond implies formation of π and σ combinations with p-like orbitals of the acceptor states perpendicular and parallel to the dimer axis, respectively. The bonding combinations are lower in energy than the levels of isolated dopants, and in the case of the considered acceptors they are degenerate with the continuum of the valence band of ZnO, while the antibonding states of pairs (denoted by stars) are higher in energy than those of isolated acceptors. (In the wurtzite structure, the picture is somewhat more complex due to the small crystal field splittings of levels, which is small and neglected here.) Figure 1c shows the calculated levels of Ag-N. As it follows from the figure, the level order is typical for a molecule. In the case of a neutral pair, the π* triplet is fully occupied with 4 electrons, and the σ* singlet is higher in energy and empty.

Finally, comparing Ag-Ag with Ag-N one can see that in the former case the distance between the dopants is larger, the bonding-antibonding splitting is smaller, and $E_{bind}$ is lower. This is full in agreement with the molecular-like picture. In all cases, $E_{bind}$ is consistent with the value of the bonding-antibonding splitting and the occupation of the acceptor orbitals. For example, the energy gain from the changes of eigenenergies is about 0.4 eV for an Ag-N pair, explaining most of $E_{bind}$ = 0.6 eV. Finally, one can observe that formation of donor-donor pairs can be driven by the same molecular-like mechanism, but the binding energy is expected to be lower since the donor impurity states in the gap are more shallow than those of the acceptor levels. Similarly, the short-range character of the coupling stems from the localization of acceptor wave functions.

## C. Binding of charged acceptor-acceptor pairs

As it was pointed out above, the molecular-like model predicts the binding energy of an acceptor pair to depend on the occupation of the molecular levels by electrons. We thus turn to the case of negatively charged acceptors, where the presence of the additional electron(s) is due to donors such as compensating oxygen vacancies or H. Donors are assumed to be distant from both Ag and N. As it follows from the molecular model, the occupation of the σ* state by an additional electron is expected to lower $E_{bind}$. This is indeed the case, since for a negatively charged (Ag-N)$^{1-}$ pair $E_{bind}$ decreases to 0.3 eV, but is still positive. The situation is different for a doubly charged pair, with two additional electrons on the σ* state. In this case the mechanism of covalent bonding is not operative, because the numbers of electrons on the bonding and antibonding states are the same. Moreover, there is also the Coulomb repulsion between the acceptors, which prevents formation of pairs. (Note that the Coulomb repulsion vanishes for the q=1- charge state). Accordingly, the calculated $E_{bind}$ of (Ag-N)$^{2-}$ is reduced to 0.05 eV. The calculated energy levels are shown in F. 1d, and with the increasing occupation they rise in energy due to the increasing electron-electron coupling.

## D. N-Ag-N triangles

The configuration and energy levels of an N-Ag-N triangle is shown in Figs. 1 b and d, respectively. The binding energy is 1.29 eV, which is correlated with a strong upward shift of the antibonding states, indicative of the corresponding downward shift of the bonding states. The same $E_{bind}$=1.3 eV is obtained for the Ag-N-Ag triangle, because the electronic structures of Ag and N are very similar. This confirms that the molecular-like mechanism of binding is largely independent of the acceptor. The highest empty antibonding state of the triangle at 0.62 eV is much higher than σ* of the Ag-N pair, 0.34 eV (Figs. 1c and 1d). As in the case of pairs, binding of negatively charged triangles is strongly suppressed, e.g., we find $E_{bind}$=0.20 eV for q=3- charge state.

(In the supercell method, when the aggregates are charged, strong interactions between the charge images can distort the final results and $E_{bind}$).

## IV. THE IMPACT OF HYDROGEN
### A. Ag-N complex with one H ion

Since hydrogen is a mobile donor, one can expect that it not only compensates intentional acceptors but also forms donor-acceptor complexes[31] and plays an active role in the aggregation of acceptors. In this Section we investigate this effect.

We first analyze binding of an Ag-N pair in the presence of one H atom. Figure 2 shows three configurations of an Ag-N pair with one H, namely (a) a N-H pair with a distant Ag, (b) an Ag-H pair with a distant N, and (c) a nearest neighbor N-Ag pair binding H, which are denoted in Fig. 2 as Ag-NH, (AgH-N), and (AgNH), respectively. In intrinsic ZnO, the equilibrium site of $H^+$ is within the Zn-O bond close to the negatively charged anion.[29,31] Similarly to the case of H in pure ZnO,[30] H in ZnO:(Ag,N) is located at the bond center between the acceptor and its neighbor,[40] and it induces a large displacement of its cation, leaving the anion only slightly displaced. In fact, in the case of an N-H pair the Zn neighbor of H is displaced from equilibrium (Fig. 2a) and N is almost non-displaced. In the case of Ag-H (Fig. 2b), the large shift of Ag brakes the axial symmetry, and both H and Ag are displaced from symmetric sites, which can explain the observations of Ref. [9]. We also find that hydrogen binds preferentially to N, and the energy of N-H is lower by 0.2 eV than that of the Ag-H configuration. This is because the local distortions are larger, and the elastic strain is higher, for Ag-H. Finally, in the configuration of the N-H-Ag nearest neighbor complex (Fig. 2c) both features are present, since Ag is off-site, N is on-site, and H assumes an almost interstitial location.

Given that an additional electron weakens binding of $(Ag-N)^{1-}$, one could expect that the presence of H donor lowers $E_{bind}$ (here, $E_{bind}$ is the energy of the configurations Fig.2b and Fig.2c relative to that of Fig. 2a). However, the opposite effect takes place, since the calculated $E_{bind}$=0.7 eV, which is higher not only than $E_{bind}$=0.3 eV of $(Ag-N)^{1-}$, but also than $E_{bind}$=0.6 eV of a neutral Ag-N pair. The increased binding is due to the attractive and localized potential of the proton. Moreover, the proton potential also strongly affects the energy levels of Ag-N. In particular, the energy of the singly occupied σ* level of $(Ag-N)^{1-}$ is 0.64 eV. The close proximity of $H^+$ lowers this energy to 0.14 eV, Fig. 2d. The mechanism making a triangle acceptor-donor-acceptor more shallow than the isolated acceptor was discussed in Refs. [33,41]. This effect favors formation of the Ag-H-N complex, since it overcompensates the weakening of the Ag-N bond by the additional electron. Finally, we find that the binding energy of $H^+$ by the Ag-N nearest neighbor pair is 1.2 eV, which is large relative to the growth and anneal temperatures.

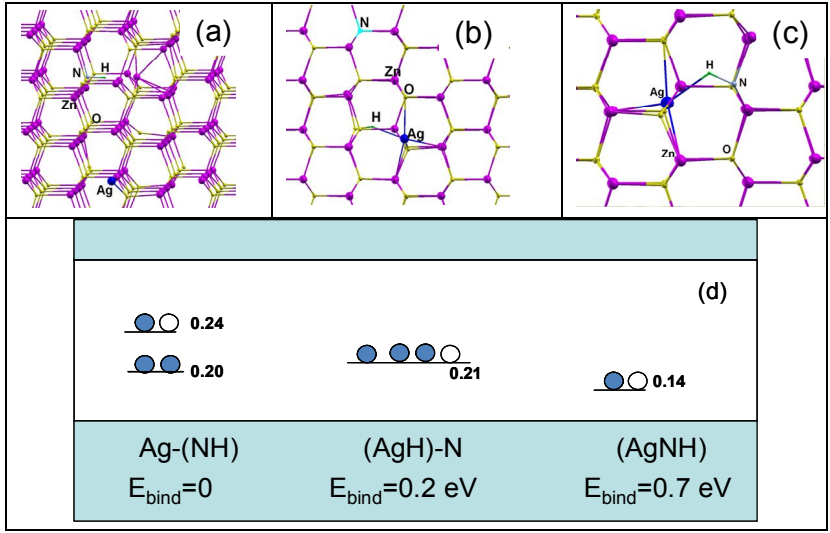

FIG. 2. (Color online) Atomic configurations and energy levels of Ag-N-H complexes: (a) N-H pair with a remote Ag, (b) Ag-H pair with a remote N, and (c) the Ag-H-N nearest neighbor complex. The corresponding energy levels (in eV) are shown in the lower part of the Figure. Magenta, yellow, big navy blue and small pale blue balls represent Zn, O, Ag and N atoms, respectively.

### B. Ag-N complex with two H ions

Contrary to the case of the compensation by a remote "generic" donor discussed in Sec. 3, formation of an Ag-N pair is not blocked even in the case of its compensation by two H ions. The configuration of distant Ag and N, each decorated with H ion, is shown in Fig. 3a, while the ground state configuration of Ag-N with two $H^+$ lower by 0.5 eV is shown in Fig. 3b. In the latter case, Ag is strongly displaced from the ideal site, and the two protons are close to N and seem to form a $H_2$ molecule. However, this is not the case, since electrons from H are transferred to N and Ag, and $H^+$ ions repel each other. This is clearly reflected by the fact that the distance between the $H^+$ ions in Fig. 3b, 2.02 Å, is almost three times longer than the H-H distance in the $H_2$ molecule in vacuum, 0.74 Å. The ground state configuration is stabilized by the Coulomb interactions between oppositely charged ions. Considering the electronic structure, the close proximity of the two $H^+$ lowers the levels of $(Ag-N)^{2-}$ from 0.74 eV (Fig. 1d) to 0.21 eV (Fig. 3c), as it was the case of Ag-H-N.

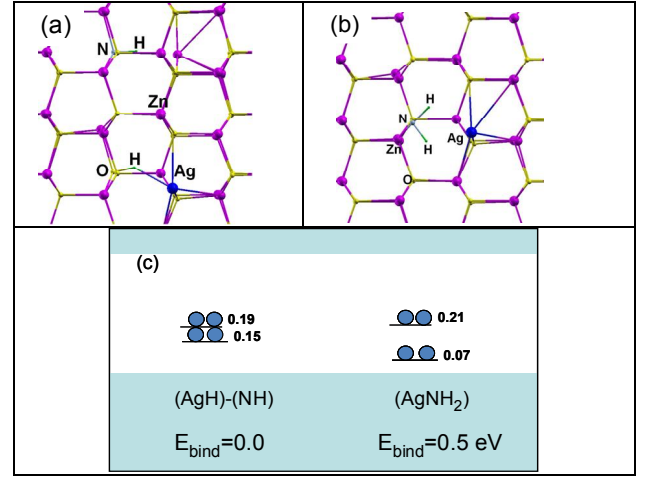

FIG. 3. (Color online) Atomic configurations of Ag-N with two H ions. (a) distant Ag-H and N-H, and (b) the equilibrium configuration. Lower panel shows the corresponding energy levels. Magenta, yellow, big navy blue and small pale blue balls represent Zn, O, Ag and N atoms, respectively.

### C. N-Ag-N triangle with H

Finally, we briefly discuss the complex of an N-Ag-N triangle with H. The configuration of distant acceptors and that of the ground state are shown in Fig. 4. The presence of $H^+$ lowers the binding energy of the triangle from 1.29 eV (Fig. 1) to 0.9 eV, and this energy is still substantial. In other words, like in the case of Ag-N, hydrogen does not prevent formation of Ag-N complexes. On the other hand we note that the energy of the acceptor level of Ag-N-Ag is lowered from 0.62 eV (Fig. 1) to 0.21 eV by the attractive potential of $H^+$.

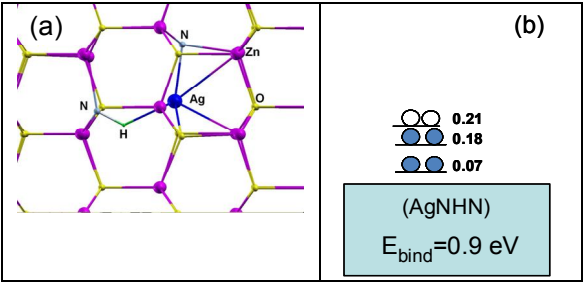

FIG. 4. (Color online) N-Ag-N nearest neighbor triangle with one H, and the corresponding energy levels. Magenta, yellow, big navy blue and small pale blue balls represent Zn, O, Ag and N atoms, respectively.

## V. IMPACT OF COMPLEXES OF Ag, N, AND H ON THE ELECTRONIC STRUCTURE AND DOPING EFFICIENCY

We now summarize the effect of formation of complexes on the electronic structure and doping efficiency. According to the results shown in Fig. 1, the calculated impurity levels of Ag and N are about 0.17 eV, *i.e.*, they are very close and relatively shallow. The nearest-neighbor Ag-N pair acts as a double acceptor with the impurity level at $E_{imp}$=0.34 eV, which is deeper than $E_{imp}$ of isolated acceptors. Thus, formation of acceptor-acceptor pairs considerably lowers doping efficiency. Pairing also occurs for both Ag and N monodoping. However, such pairs are less bound than Ag-N, and in this case the resulting double acceptor level is more shallow. From this point of view monodoping is more efficient. Turning to N-Ag-N triangles, its lowest half-empty state is at 0.27 eV (Fig. 1d), *i.e.*, it is lower than that of an Ag-N pair (0.34 eV). This feature is beneficial for doping efficiency, however one needs three acceptors to get a triangle. To make the comparison between various configurations more quantitative we observe that doping efficiency is given in particular by the Boltzmann factor, exp($-E_{imp}/k_BT$), where $k_B$ is the Boltzmann constant, and $T$ is the temperature. At $T$=300 K, using the above values one finds that formation of triangles and pairs lowers the efficiency by one and two orders of magnitude, respectively, relative to the case of isolated acceptors.

In the presence of H in ZnO, formation of complexes of Ag-N pairs with H lowers $E_{imp}$, see Fig. 2d. In particular, the Ag-N-H nearest-neighbor complex is a single acceptor with $E_{imp}$=0.14 eV, which is lower than $E_{imp}$ of isolated Ag and N. As explained, this is due to the attractive potential of the proton. Thus, when the H concentration is one half of the acceptor concentration or less, formation of Ag-N-H should not affect strongly the conductivity of the ZnO layer. For higher H concentrations the compensation of Ag and/or N acceptors by H donors takes place, and can eventually lead to fully compensated, insulating ZnO samples.

In the limit of low impurity concentrations, both acceptors are relatively shallow. However, doping ZnO requires high acceptor concentrations. In this situation, the impurity band forms, which is a superposition of levels of various configurations. This band is wide, with impurity states extending up to 0.7 eV. Broadening is due not only to the overlap of impurity wave functions, but also to the formation of pairs and triangles. The non-uniform impurity distribution occurs because the statistics is affected by finite binding energies.

## VI. CONCLUSIONS

In summary, efficiency of dual doping of ZnO with Ag and N acceptors was investigated by first principles calculations. Formation of few atom Ag-N complexes was analyzed, and the impact of the possible presence of H in ZnO was taken into account. The acceptor levels of isolated Ag and N are found to be shallow and very close. However, Ag and N have a tendency to form nearest-neighbor pairs and triangles, with binding energies of about 0.5 eV. Formation of such complexes increases acceptor energies, and thus lowers the doping efficiency.

A molecular-like model of the acceptor pair formation is put forward, in which the proximity of two acceptors induces formation of bonding and antibonding combinations of their acceptor levels. This explains the calculated features characterizing Ag-N pairs and triangles. In particular, the binding energy of nearest neighbor Ag-N pairs, 0.7 eV, is higher than that of an Ag-Ag or N-N pair due to the much smaller acceptor-acceptor distance. This is also reflected in the stronger bonding-antibonding splitting of molecular-like states of acceptor levels. Moreover, the presence of "generic" donors in ZnO (*e.g.*, oxygen vacancies) leads to the occupation of the antibonding states by electrons, which weakens the bonding. Finally, this picture explains the tendency to form acceptor-acceptor pairs found in a variety of semiconductors.

H atoms in ZnO influence the acceptor pairing process. In contrast to remote "generic" donors that prevent the pairing, H decorates both isolated acceptors and Ag-N, and promotes formation of complexes. Next, H strongly affects their electronic structure by lowering the acceptor energies. In particular, the acceptor level of Ag-H-N is lower than $E_{imp}$ of isolated Ag or N. From the obtained results it follows that H in ZnO:Ag or ZnO:N samples is more difficult to be annealed out than in pure ZnO because of the formation of pairs and triangles with acceptors.

## ACKNOWLEDGEMENTS

The work was supported by the European Union within European Regional Development Fund through grant Innovative Economy (POIG.01.03.01-00-159/08, "InTechFun").